\documentclass[conference]{IEEEtran}
\IEEEoverridecommandlockouts
\usepackage{cite}
\usepackage{amsmath,amssymb,amsfonts}
\usepackage{algorithmic}
\usepackage{algorithm}
\usepackage{graphicx}
\usepackage{textcomp}
\usepackage{xcolor}
\usepackage{epsfig}
\usepackage{subfigure}
\usepackage{psfrag}
\def\BibTeX{{\rm B\kern-.05em{\sc i\kern-.025em b}\kern-.08em
    T\kern-.1667em\lower.7ex\hbox{E}\kern-.125emX}}
\begin{document}

\title{Energy-Efficient Proactive Caching for Fog Computing with Correlated Task Arrivals}
\author{\IEEEauthorblockN{Hong Xing${}^\ast$, Jingjing Cui${}^\S$, Yansha Deng{}$^\dag$, and Arumugam Nallanathan${}^\S$}\\
	\vspace{-.1in}
	\IEEEauthorblockA{${}^\ast$College of Information Engineering, Shenzhen University, Shenzhen, China\\
		${}^\S$School of EECS, Queen Mary University of London, London, U.K.\\
		${}^\dag$Department of Informatics, King's College London, U.K.\\
		E-mails:~hong.xing@szu.edu.cn,~j.cui@qmul.ac.uk,~yansha.deng@kcl.ac.uk,~nallanathan@ieee.org}}


\maketitle

\setlength{\baselineskip}{1.0\baselineskip}
\newtheorem{definition}{\underline{Definition}}[section]
\newtheorem{fact}{Fact}
\newtheorem{assumption}{Assumption}
\newtheorem{theorem}{\underline{Theorem}}[section]
\newtheorem{lemma}{\underline{Lemma}}[section]
\newtheorem{corollary}{\underline{Corollary}}[section]
\newtheorem{proposition}{\underline{Proposition}}[section]
\newtheorem{example}{\underline{Example}}[section]
\newtheorem{remark}{\underline{Remark}}[section]
\newcommand{\mv}[1]{\mbox{\boldmath{$ #1 $}}}
\newcommand{\mb}[1]{\mathbb{#1}}
\newcommand{\Myfrac}[2]{\ensuremath{#1\mathord{\left./\right.\kern-\nulldelimiterspace}#2}}

\begin{abstract}
With the proliferation of latency-critical applications, fog-radio network (FRAN) has been envisioned as a paradigm shift 
enabling distributed deployment of cloud-clone facilities at the network edge. In this paper, we consider proactive caching for a one-user one-access point (AP) fog computing system over a finite time horizon, in which consecutive tasks of the same type of application are temporarily correlated. Under the assumption of predicable length of the task-input bits, we formulate a long-term weighted-sum energy minimization problem with three-slot correlation to jointly optimize computation offloading policies and caching decisions subject to stringent per-slot deadline constraints. The formulated problem is hard to solve due to the mixed-integer non-convexity. To tackle this challenge, first, we assume that task-related information are perfectly known {\em a priori}, and provide offline solution leveraging the technique of semi-definite relaxation (SDR), thereby serving as theoretical upper bound. Next, based on the offline solution, we propose a sliding-window based online algorithm under arbitrarily distributed prediction error. Finally, the advantage of computation caching as well the proposed algorithm is verified by numerical examples by comparison with several benchmarks.
\end{abstract}


\section{Introduction}
Unprecedented growth of computation-extensive services (such as video streaming analysis, virtual reality (VR), and autonomous driving) prohibits the cloud-radio access network (CRAN) from continuously satisfying their latency-critical demands due to increasing transmission delay over long distance between the cloud and the users. To resolve such challenges, {\em fog-radio access network (FRAN)}, as an evolution of CRAN, is paving its way to provide ultra-reliable and low-latency (uRLLC) services for future wireless networks by pushing cloud-like capabilities, namely, {\em fog computing} and {\em edge caching}, to the network edge  \cite{Simeone2016Mag,Shih2017network}.

Fog computing, also known as {\em mobile edge computing (MEC)}, endows the edge access points (APs) with computing and storage capacities, such that low-power wireless devices can seek nearby APs that are integrated with edge servers for task offloading, thus enabling energy-saving computation in real time. In the literature, a large amount of efforts have been devoted to achieving satisfied trade-offs between the cost of the network and latency by joint management of computation and communication resource as well as task offloading decisions (see e.g., \cite{Wang2018WP,Chen18MCS,mao16dynamic}). 

Meanwhile, edge caching allows users to fetch popular contents from near by APs and/or users, thus alleviating the growing over-the-air traffic. Existing works have mainly focused on improving the efficiency of cache-enabled content distribution (see \cite{bennis14caching} and the references therein),
whereas, caching aimed for saving the edge servers from repeated computing is less studied. The authors in \cite{elbamby17proacitve}  investigated proactive caching for achieving uRLLC in fog networks. However, they assumed that the popular computation tasks that had been cached {\em a priori} can be completely reused when requested later, which is too ideal in practice, since unlike content distribution, computation services usually adopt one-time data sets that are hardly rendered the same later on. Hence, it is crucial to understand what to cache by carefully exploiting the intrinsic data correlation among task arrivals. 
Note that although \cite{xu18infocom} and \cite{hao2018access} considered joint service caching and task offloading,  they  
did not model how the computation  offloading can benefit from dynamic caching of correlated (not necessarily the same) task results. 

In this work, we study proactive caching for a fog computing system consisting of one user terminal (UT) and one AP over a finite time horizon leveraging the correlation among {\em delay sensitive} task sequence such that the task results cached at the current slot can facilitate future computing. To our best knowledge, this is the first work aimed for minimizing the long-term weighted-sum energy by jointly optimizing computation offloading policies and caching decisions. With the correlation lying among three consecutive slots and imperfect task-input prediction, first, we provide an offline solution based on semi-definite relaxation (SDR), which serves as an performance upper bound. Next, we propose a sliding-window inspired online solution taking causally known prediction error into account. Finally, numerical results show striking performance gains brought by computation caching as well as the effectiveness of the proposed online algorithm. 

We use the upper case boldface letters for matrices and lower case boldface ones for vectors. The superscripts \((\cdot)^T\) and \((\cdot)^\ast\) represent, respectively, the transpose and the optimum solution of vectors or matrices. We also denote the trace of a matrix by \({\sf Tr}(\cdot)\). 

\section{System Model and Problem Formulation}\label{sec:System Model and Problem Formulation}
\begin{figure}[htp]
	\centering
	\includegraphics[width=2.5in]{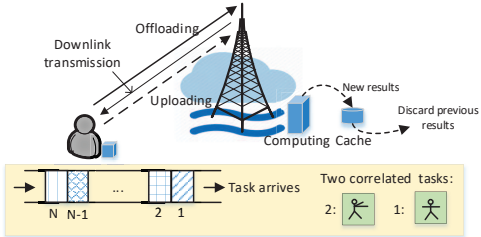}
	\caption{System model of the one-user one-server fog computing system.}\label{fig:system model}
	\vspace{-0.2in}
\end{figure}

We consider a fog computing system consisting of one UT equipped with one single antenna, and one access point (AP) equipped with $M$ antennas, an edge server and cache facilities.  During slot $i$, the UT solicits the nearby AP for computation task offloading. In this paper, we focus on a finite slotted-time horizon with each slot lasting $T$ seconds, denoted by \(\mathcal{N}=\{1,\ldots,N\}\), over which sequential tasks featuring temporally correlated input data arrive at the UT as shown in Fig. \ref{fig:system model}. We assume that each task has to be executed by the end of the time slot. Since the computing results obtained at the current slot are also correlated with those at future slots, current task results can be cached at the AP to facilitate the future computation\footnote{A typical example is matrix-vector multiplication of \(\mv y_i=\mv A\mv x_i\), \(i\in\mathcal{N}\), where \(x_i\)'s is the encoded task-input data. Supposing \(\mv x_{i}=\mv x_{i-1}+\mv\varepsilon\) with a sparse error \(\mv\varepsilon\), the current computation can benefit from caching at slot $i-1$ by executing only \(\mv\varepsilon\) with much shorter input length.\label{example}}. Due to the extra overhead caused by caching (delay, energy, storage), it may not be optimal to cache all the execution results at the edge server. Therefore, we introduce the following variable \(I_i\), \(i\in\mathcal{N}\), to indicate whether the AP decides to cache the results at the end of slot $i$:
\begin{align}
I_i=\begin{cases}
1,\, \mbox{if the BS decides to cache the results, }\\ 
0,\, \mbox{otherwise.}
\end{cases}\label{eq:caching indicator}
\end{align} 

As a result, the workflow of the cache-enabled fog computing system in consideration can be described as follows. The UT offloads a proportion of the task to the AP while performing local computing for the rest of the task. If the AP decides not to cache the task results of the current slot, the UT just need to receive the execution results from the AP, otherwise the UT is also required to upload its local computing results to the AP at the end of the current slot.
Since the AP is usually of sufficient communications resource. e.g., high transmitting power, we ignore the delay/energy caused by results downloading at the UT in the sequel. 

\subsection{Local Execution, Task Offloading and Computation Uploading at the UT}
We assume that the length of task-input bits at slot \(i\in\mathcal{N}\), denoted by \(L_i\)'s, is predictable but with finite estimation error shown as  \(L_i=\hat L_i+\Delta L_i\), in which \(\{\Delta L_i\}\) can be an arbitrary (deterministic or stochastic) sequence.  At any slot \(i\in\mathcal{N}\), the exact task-input length up to slot $i$, i.e., \(L_k\)'s for \(k\le i\), is  known to the AP, while only the predicted task-input length, i.e., \(\hat L_k\)'s for \(k>i\), is available for all future slots. We model the task-input bits that are required to be executed at slot $i$ in terms of previous caching decisions as follows:
\begin{multline}
D_i=L_i\Big((I_{i-1}\tau_1+\ldots+\prod_{j=1}^{k-1}(1-I_{i-j})I_{i-k}\tau_k+\ldots\\
+\prod_{j=1}^{r-1}(1-I_{i-j})I_{i-r}\tau_r+\prod_{j=1}^{r}(1-I_{i-j})\Big), \label{eq:re-executed input}
\end{multline} where \(\mv\tau=[\tau_1,\ldots\tau_r]^T\) with increasing \(\tau_j\in[0,1]\), \(j=1,\ldots,r\), is a prescribed vector capturing the diminishing effect of the  previously cached  results on reducing the current task-input length. Note that only the latest cached results are useful. (E.g., if \(I_{i-1}=0\) and \(I_{i-2}=1\), \eqref{eq:re-executed input} reduces to \(D_i=L_i\tau_2\) in spite of the values that \(I_{i-k}\) for \(k\ge 3\)  take.) In addition,  any results cached far more than $r$ slots before are assumed to be no longer exploitable.

{\bf Local Execution} The cache-enabled task-input bits \(D_i\)'s will be divided into \(l_i\) and \(D_i-l_i\) for local and remote execution, respectively, where \(l_i\in[0,D_i]\). The required CPU cycles for UT's local execution  at slot $i$ is given by \(c_{\rm loc}l_i\) \cite{Wang2018WP}, where \(c_{\rm loc}\) in cycles per bit depends on the application type and the CPU architecture of the UT. Assuming constant CPU frequency \(f_{\rm loc}\) adopted by the UT, the corresponding energy consumption for local computation at slot $i$ is expressed as \cite{Mach17survey}
\begin{align}
E_{c,i}^{\rm loc}=\kappa_{\rm loc}c_{\rm loc}l_if_{\rm loc}^2,\label{eq:local computation energy}
\end{align} where \(\kappa_{\rm loc}\) is the effective capacitance coefficient of the UT's CPU chip. 

{\bf Task Offloading}  By applying maximum ratio combing (MRC) at the AP, the achievable offloading rate at slot $i$,   \(i\in\mathcal{N}\), is thus given by \(r_i^{\rm off}=B_{\rm off}\log_2(1+p_ih_i)\), where \(h_i\) is the normalized channel gain from the UT to the AP at slot $i$, and \(B_{\rm off}\) is the pre-assigned transmission bandwidth (BW) for task offloading\footnote{We assume that frequency division multiple access (FDMA) is adopted for task offloading and computation results uploading, respectively.}. It thus takes \(t_i^{\rm off}=\Myfrac{(D_i-l_i)}{r_i^{\rm off}}\), for task offloading, and the associated  energy consumption for task offloading is given by
\begin{align}
E^{\rm off}_i=\frac{p_i(D_i-l_i)}{r_i^{\rm off}}. \label{eq:task offloading energy}
\end{align} 

{\bf Computation Uploading} Suppose that there is little cache capacity allocated for computation caching at the UT. When the current task results are decided to be cached at the end of slot $i$,  \(i\in\mathcal{N}\), the UT needs to upload its locally executed results to the AP so as to maintain the integrity of computation results for future use. Given the UT's uploading rate, \(r_i^{\rm up}\)'s, the consumed energy for computation uploading at the UT is thus given by 
\begin{align}
E^{\rm up}_i=I_i\frac{p_iR_i}{r_i^{\rm up}}, \label{eq:computation uploading energy}
\end{align} 
where \(R_i\)  is the length of the task-output bits, which is assumed to have been perfectly profiled given the type of application (c.f. footnote \ref{example}).

\subsection{Remote Execution and Computation Caching at AP}
In the considered model, the AP is responsible for profiling the task information (\(\hat L_i\)'s and \(R_i\)'s) as well as the channel state information (CSI) (\(h_i\)'s \(g_i\)'s), and collecting other required information {a priori}. Based on these information, the AP will dynamically make and inform the UT of the  caching decisions and the offloading policies.

{\bf Remote Execution} Similar to \eqref{eq:local computation energy}, the energy consumption for remote execution is expressed as
\begin{align}
E_{c,i}^{\rm e}=\kappa_{\rm e}c_{\rm e}(D_i-l_i)f_{\rm e}^2,\label{eq:remote computation energy}
\end{align} where \(\kappa_{\rm e}\) and \(c_{\rm e}\) denote the effective capacitance coefficient, and the number of cycles required for executing one task-input bit at the edge server's CPU, respectively. 

{\bf Computation Caching} If the AP decides to cache the task results at the current slot, it then expects to receive UT's uploading of  its local computing results before combining them with the remotely executed task to form an integrated copy of the task results ready for caching. 

\subsection{Problem Formulation}
We are interested in minimizing the total weighted-sum energy consumption over the finite horizon \(\mathcal{N}\), i.e., \(\sum_{i\in\mathcal{N}}(\alpha_1(E_{c,i}^{\rm loc}+E^{\rm off}_i+E^{\rm up}_i)+\alpha_0E_{c,i}^{\rm e})\), where \(\alpha_1\) and \(\alpha_0\) satisfying \(\alpha_1+\alpha_0=1\) are the coefficients balancing  the energy saving priority between the UT and the AP. Under the per-slot deadline constraint for each task, we aim to jointly optimize the computation offloading policies \(\{l_i\}\) and the binary caching decisions \(\{I_i\}\). Combining \eqref{eq:local computation energy}, \eqref{eq:task offloading energy}, \eqref{eq:computation uploading energy}, and \eqref{eq:remote computation energy}, the long-term energy minimization problem is thus formulated as:
\begin{subequations}
	\begin{align}
	\mathrm{(P1)}:&~\mathop{\mathtt{Min}}_{\{l_i,I_i\}}~\sum\limits_{i\in\mathcal{N}}\bigg(\alpha_1\bigg(\kappa_{\rm loc}c_{\rm loc}l_if_{\rm loc}^2+I_i\frac{p_iR_i}{r_i^{\rm up}}\notag\\
	&~+\frac{p_i(D_i-l_i)}{r_i^{\rm off}}\bigg)+\alpha_0\kappa_{\rm e}c_{\rm e}(D_i-l_i)f_{\rm e}^2\bigg)\notag\\
	\mathtt {s.t.}&~\frac{D_i-l_i}{r^{\rm off}_i}+\frac{c_{\rm e}(D_i-l_i)}{f_{\rm e}}\le T,\; \forall i\in\mathcal{N},\label{C:remote computation deadline}\\
	&~\frac{c_{\rm loc}l_i}{f_{\rm loc}}+I_i\frac{R_i}{r^{\rm up}_i}\le T,\; \forall i\in\mathcal{N},\label{C:local compuation deadline}\\
	&~0\le l_i\le D_i,\; \forall i\in\mathcal{N},\\ 
	&~I_i\in\{0,1\},\; \forall i\in\mathcal{N}.\label{C:binary constraints}
	\end{align} 
\end{subequations}

\section{Offline Computation Offloading and Caching}
In this section, we consider offline solution for problem (P1) by assuming that the predictable task-input length \(\{L_i\}\)  are perfectly known {\em a priori} at the AP. The offline solution thus serves as fundamental performance upper bound for all other online schemes that are designed for practical implementation. In this paper,  we focus on a special case of $r=2$ (c.f.~\eqref{eq:re-executed input}). More general cases will be studied in our future work.

The major difficulty for solving (P1) lies in the binary variables \(I_i\)'s. To tackle this challenge, first, we equivalently formulate \eqref{C:binary constraints} as \(I_i(I_i-1)=0\), \(\forall i\in\mathcal{N}\)), and  then transform the  problem into a  quadratically constrained quadratic program (QCQP) in terms of \(\mv I=[I_1,\ldots,I_N]^T\). Next, we convert the QCQP into a semi-definite programming (SDP) as follows.
First, we define \(\mv F^\prime=[(1-\tau_2)\mv F, \tfrac{1}{2}\mv v; \tfrac{1}{2}\mv v^T, 0]\), where \(\mv F=\sum\limits_{i=3}^N\tfrac{p_iL_i}{r_i^{\rm off}}\mv G_{i-2,i-1}\), \(\mv G_{i-2,i-1}\) is a symmetric matrix with only \(\mv G_{i-2,i-1}(i-2,i-1)\) and \(\mv G_{i-2,i-1}(i-1,i-2)\) being $\tfrac{1}{2}$, \(\mv v=(\tau_1-1)\sum\limits_{i=2}^N\tfrac{p_iL_i}{r_i^{\rm off}}\mv e_{i-1}+(\tau_2-1)\sum\limits_{i=3}^N\tfrac{p_iL_i}{r_i^{\rm off}}\mv e_{i-2}\), and \(\mv e_j\) denotes a vector with only the $j$th element being $1$; \(\mv W=[\mv 0_{N\times N}, \tfrac{1}{2}\mv w; \tfrac{1}{2}\mv w^T, 0]\), where \(\mv w=[\tfrac{p_1R_1}{r_1^{\rm up}},\ldots,\tfrac{p_NR_N}{r_N^{\rm up}}]^T\);  \(\mv u=[\tfrac{p_1}{r_1^{\rm off}},\ldots,\tfrac{p_N}{r_N^{\rm off}}]^T\), \(\mv G^\prime=[(1-\tau_2)\mv G, \tfrac{1}{2}\mv s; \tfrac{1}{2}\mv s^T, 0]\), where \(\mv G=c_{\rm e}\sum\limits_{i=3}^NL_i\mv G_{i-2,i-1}\), and \(\mv s=(\tau_1-1)c_{\rm e}\sum\limits_{i=2}^NL_i\mv e_{i-1}+(\tau_2-1)c_{\rm e}\sum\limits_{i=3}^NL_i\mv e_{i-2}\);  \(\mv E_i=[\mv 0_{N\times N} ,\tfrac{1}{2}\mv e_i; \tfrac{1}{2}\mv e_i^T, 0]\); and \(\mv U_i=[\mathrm{diag}(\mv e_i), -\tfrac{1}{2}\mv e_i; -\tfrac{1}{2}\mv e_i^T, 0]\). Next, we introduce \(\mv a=[\mv I; 1]\) and \(\mv A=\mv a\mv a^T\). Then, by relaxing the rank-one constraint for \(\mv A\) \cite{Luo10SDR} and  some manipulations, problem (P1) is recast into an SDP as shown in the following  proposition.
\begin{proposition}
	By relaxing the rank-one constraint, problem (P1) is equivalent to an SDP shown below:
	\begin{subequations}
		\begin{align}
		\mathrm{(P1^\prime)}:&~\mathop{\mathtt{Min}}_{\mv A, \mv l}~\alpha_1\left.({\sf Tr}(\mv A\mv F^\prime)+{\sf Tr}(\mv A\mv W)-\mv u^T\mv l+\right.\notag\\
		&~\left.\kappa_{\rm loc}c_{\rm loc}f_{\rm loc}^2\mv 1^T\mv l\right)+\alpha_0\kappa_{\rm e}f_{\rm e}^2\left({\sf Tr}(\mv A\mv G^\prime)-c_{\rm e}\mv 1^T\mv l\right)\notag\\
		\mathtt {s.t.}&~\big(\frac{1}{r_i^{\rm off}}+\frac{c_{\rm e}}{f_{\rm e}}\big)(D_i(\mv A)-\mv e_i^T\mv l)\le T,\; \forall i\in\mathcal{N},\\
		&~\frac{R_i}{r_i^{\rm up}}{\sf Tr}(\mv A\mv E_i)+\frac{c_{\rm loc}}{f_{\rm loc}}\mv e_i^T\mv l\le T,\; \forall i\in\mathcal{N},\\
		&~\mv e_i^T\mv l-D_i(\mv A)\le 0,\; \forall i\in\mathcal{N},\label{C:non-negative computation offloading}\\
		&~{\sf Tr}(\mv A\mv U_i)=0,\; \forall i\in\mathcal{N},\\
		&~\mv A(N+1,N+1)=1,\\
		&~\mv l\ge 0,\,\, \mv A\succeq 0.
		\end{align} 
	\end{subequations} \label{prop:SDP transformation}
\end{proposition}
\begin{IEEEproof}
Due to the space limitation, we only provide a key step in the proof, i.e., to express \(D_i\)'s in terms of \(\mv A\). Since \(D_i=L_i((\tau_1-1)I_{i-1}+(\tau_2-1)I_{i-2}+(1-\tau_2)I_{i-1}I_{i-2}+1)\), \(i\ge 3\), it follows that  \(D_i=L_i{\sf Tr}(\mv A\mv H_i)\), \(i\ge 3\),  where \(\mv H_i=[(1-\tau_2)\mv G_{i-2,i-1}, \tfrac{1}{2}((\tau_1-1)\mv e_{i-1}+(\tau_2-1)\mv e_{i-2}); \tfrac{1}{2}((\tau_1-1)\mv e_{i-1}+(\tau_2-1)\mv e_{i-2})^T,1]\).
\end{IEEEproof}

As \(\mathrm{(P1^\prime)}\) is an SDP, we can solve \(\mathrm{(P1^\prime)}\) by some off-the-shelf convex software tools, such as CVX\cite{CVX}. Since there is no guarantee that \(\mv A^\ast\) for  \(\mathrm{(P1^\prime)}\) is rank-one, it in general only serves as a lower-bound solution for (P1). To construct the binary caching decisions, we need to retrieve \(\mv I\) from \(\mv A^\ast\). Specifically, if \({\rm rank}(\mv A^\ast)=1\), \(\mv I^\ast\) can be recovered by singular-value decomposition (SVD) such that \(\mv A^\ast=\mv a^\ast\mv a^{\ast T}\). Otherwise,  we propose to approximate \(I_i\)'s as follows.
\begin{align}
I_i^{\rm app}={\rm round}(A^\ast(i,N+1)),\; i\in\mathcal{N},\label{eq:approximate caching decisions}
\end{align} which is based on the following lemma \cite{Chen18MCS}.
\begin{lemma}
The optimum \(\mv A^\ast\) for problem \(\mathrm{(P1^\prime)}\) satisfies \(\mv A^\ast(i,N+1)\in[0,1]\), \(i\in\mathcal{N}\).\label{lemma: value range for the last col of A}
\end{lemma}
Once \(\mv I^{\rm app}\) is ready, the corresponding offloading policies \(\mv l^{\rm app}\) can be easily obtained by solving \(\mathrm{(P1^\prime)}\) with  \(\mv A=\mv a^{\rm app}\mv a^{{\rm app}T}\) fixed  (\(a^{\rm app}=[\mv I^{\rm app};1]\)), which then turns out to be a linear programming (LP) problem in terms of \(\mv l\).

As per Lemma \ref{lemma: value range for the last col of A}, when \(\mv A^\ast\) is rank-one, the approximation is tight because  \( I_i^\ast= a_i^\ast=a_i^\ast a_{N+1}^\ast=A^\ast(i,N+1)\). It thus implies that the effectiveness of the approximated caching decisions primarily depends on the rank property of \(\mv A^\ast\). The following proposition reveals a sufficient condition  for achieving low-rank \(\mv A^\ast\) that is easily satisfied in practice \cite{Chen18MCS}. 
\begin{proposition}
When the constraints given by \eqref{C:non-negative computation offloading} are all inactive, i.e.,  non-zero task offloading at all the slots,  \({\rm rank}(\mv A^\ast)\le 2\).
\end{proposition}
\begin{IEEEproof}
Please refer to Appendix \ref{appendix:proof of rank-two for A}.
\end{IEEEproof}

\section{Online Computation Offloading and Caching}
In the previous section, we have provided an SDR-based offline solution under the ideal assumption that the random task-input length \(L_i\)'s  is  perfectly predicted without error. In this section, inspired by the offline solution, we propose a sliding-window based online scheme that applies to error sequence \(\{\Delta L_1,\ldots,\Delta L_N\}\) following arbitrary stochastic process \cite{Rahbar15microgrid}. 

Specifically, as stated in Section \ref{sec:System Model and Problem Formulation}, at any slot $i$, the exact task-input  length is  perfectly known up to the current slot, i.e., \(\{L_1,\ldots,L_i\}\), whereas only the predictable task-input length, i.e., \(\hat L_{i+1}, \ldots, \hat L_N\), is available for all future slots.
First, we define a set \(\mathcal{S}=\{1,\ldots,S\}\), where $S$ is the length of the sliding-window. Note that since the parameter $S$ balances between exploitation of the long-term prediction and accuracy of the algorithm, it is required to be carefully chosen in practice.
Second, we focus on minimizing the weighted-sum energy over the span of the sliding-window from slot $i$, i.e., slots \(\{i,\ldots,i+S-1\}\). Then, by specifying the parameters using their consecutive $S$-slot values from slot $i$\footnote{When the last index of the sliding-window exceeds $N$, we substitute the (prediction) values of the parameters from slot $1$ to $S-1$ for those from slot $N+1$ to $N+S-1$.}, e.g., \(\{L_1^{(i)},L_2^{(i)},\ldots,L_S^{(i)}\}=\{L_i,\hat L_{i+1},\ldots,\hat L_{i+S-1}\}\), we sequentially solve the following problem for all the slots.
 	\begin{subequations}
	\begin{align*}
	\mathrm{(P1\text{-}ol)}:&\kern-8pt~\mathop{\mathtt{Min}}_{\mv A^{(i)}, \mv I^{(i)}}\kern-10pt~\alpha_1\big({\sf Tr}(\mv A^{(i)}\mv F^{\prime(i)})+{\sf Tr}(\mv A^{(i)}\mv W^{(i)})-\mv u^{(i)T}\mv l^{(i)}\notag\\
	&\kern-.38in+\kappa_{\rm loc}c_{\rm loc}f_{\rm loc}^2\mv 1^T\mv l^{(i)}\big)+\alpha_0\kappa_{\rm e}f_{\rm e}^2\big({\sf Tr}(\mv A^{(i)}\mv G^{\prime(i)})-c_{\rm e}\mv 1^T\mv l^{(i)}\big)\notag\\
	\mathtt {s.t.}&~\big(\frac{1}{r_j^{{\rm off}(i)}}+\frac{c_{\rm e}}{f_{\rm e}}\big)(D_j^{(i)}(\mv A^{(i)})-\mv e_j^T\mv l^{(i)})\le T,\;  \forall j\in\mathcal{S},\\
	&~\frac{R_j^{(i)}}{r_j^{{\rm up}(i)}}{\sf Tr}(\mv A^{(i)}\mv E_j)+\frac{c_{\rm loc}}{f_{\rm loc}}\mv e_j^T\mv l^{(i)}\le T,\; \forall j\in\mathcal{S},\\
	&~\mv e_j^T\mv l^{(i)}-D_j^{(i)}(\mv A^{(i)})\le 0,\; \forall j\in\mathcal{S},\\
	&~{\sf Tr}(\mv A^{(i)}\mv U_j)=0,\; \forall j\in\mathcal{S},\\
	&~\mv A^{(i)}(N+1,N+1)=1,\\
	&~\mv l^{(i)}\ge 0,\,\, \mv A^{(i)}\succeq 0,
	\end{align*} 
\end{subequations} where \(\mv e_j\), \( j\in\mathcal{S}\), is similarly defined as in \(\mathrm{(P1^\prime)}\), and so are \(\mv E_j\)'s and \(\mv U_j\)'s through proper dimension modification. Next, by reconstructing \(\mv I^{{\rm app}(i)}\)  and \(\mv l^{{\rm app}(i)}\) from the solution to \(\mathrm{(P1\text{-}ol)}\), we attain the proposed online computation offloading policies \(\{\tilde l_i\}\) and caching decisions \(\{\tilde I_i\}\) by \(\tilde l_i=l_1^{{\rm app}(i)}\) and \(\tilde I_i=I_1^{{\rm app}(i)}\), \(i\in\mathcal{N}\), respectively. The above procedure for the online scheme is summarized in Table \ref{table:online algorithm}.
\vspace{-1em}
{\small\begin{table}[hp]
		\begin{center}
			\vspace{0.5em}
			\caption{Proposed Online Algorithm for Problem (P1)}\label{table:online algorithm}
			\vspace{-0.5em}
			\hrule
			\vspace{0.75em}
			\begin{algorithmic}[1]
				\REQUIRE \(i\leftarrow 1\)
				\REPEAT
				\STATE Solve \(\mathrm{(P1\text{-}ol)}\) at  slot $i$, and obtain its optimal solution \(\mv A^{(i)\ast}\);
				\STATE Reconstruct \(\mv I^{{\rm app}(i)}\) based on \(\mv A^{(i)\ast}\) by similar means of  \eqref{eq:approximate caching decisions};
				\STATE Given \(\mv I^{{\rm app}(i)}\), solve the reduced LP associated with \(\mathrm{(P1\text{-}ol)}\) to obtain \(\mv l^{{\rm app}(i)}\);
				\STATE \(\tilde I_i\leftarrow I_1^{{\rm app}(i)}\) and \(\tilde l_i\leftarrow l_1^{{\rm app}(i)}\);
				\STATE \(i\leftarrow i+1\).
				\UNTIL \(i=N\)\\
				\ENSURE \(\{\tilde I_i, \tilde l_i\}\)
			\end{algorithmic}
			\vspace{0.75em}
			\hrule
		\end{center}
		\vspace{-2em}
	\end{table}}

\section{Numerical Results}\label{sec:Numerical Results}
In this section, we verify the effectiveness of our proposed online computation offloading and caching scheme against theoretical performance upper bound and other benchmark schemes through numerical simulations. Specifically, `Lower-bound' shows the optimal solution to   \(\mathrm{(P1^\prime)}\) based on SDR, which is only achievable when the approximation is tight; `Random caching' is obtained by setting \(\{I_i\}\) as a $\tfrac{1}{2}$-Bernoulli process; `No caching' refers to the results ignoring the correlation among task-input data; and ``All caching'' provides the case when \(\{I_i=1\}\). At each slot, we consider Rayleigh fading channel models with the distance-dependent pathloss set as $-117$dB ($0.5$km) over transmission BWs of \(B_{\rm off}=B_{\rm up}=2.5\)MHz. The estimation of the task-input length follows a uniform distribution, denoted by \(\hat L _i\sim\mathcal{U}[10^5,10^6]\)bits, \(i\in\mathcal{N}\), and the profile of the associated task-output length is set as \(R_i\sim\mathcal{U}[10^5, 10^6]\)bits. Other parameters are set as follows unless otherwise specified: \(M=3\); \(\alpha_1=0.85\), \(\alpha_0=0.15\); \(\{\tau_1=\tfrac{1}{2}, \tau_2=\tfrac{3}{4}\}\); \(\{p_i=24\}\)dBm; \(f_{\rm loc}=800\)MHz, \(f_{\rm e}=2\)GHz; \(C_{\rm loc}=C_{\rm e}=10^3\) cycles/bit; and \(\kappa_{\rm loc}=\kappa_{\rm e}=10^{-28}\). The results shown below are obtained by averaging over $500$-time realizations of the predication error sequence \(\{\Delta L_i\}\), in which \(\Delta L_i\)'s is modelled as $i.i.d.$ Gaussian variables with zero mean and variance of \(\sigma^2\).

\begin{figure}[tb]
	\centering
	\includegraphics[width=2.8in]{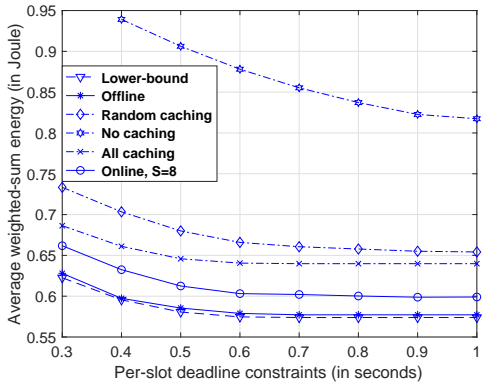}
	\caption{Average weighted-sum energy versus the per-slot deadline constraint.}\label{fig:energy vs deadline}
	\vspace{-0.2in}
\end{figure}

Fig. \ref{fig:energy vs deadline} shows the average weighted-sum energy versus the computation deadline $T$ with \(\sigma^2=10^4\). It is observed that the weighted-sum energy for all the schemes gradually goes down as the per-slot deadline gets extended, which is intuitively true, since the longer $T$ is, the higher the chances that more of the task-input bits can be executed locally within the deadline, which thus saves UT's energy for task offloading. The approximate offline solution is also shown to approach the lower-bound SDR solution with negligible gap. Furthermore, the proposed online joint computation offloading and caching scheme outperforms all the other fixed-caching schemes, which corroborates the importance of computation caching in latency-critical scenarios.

\begin{figure}[tb]
	\centering
	\includegraphics[width=2.8in]{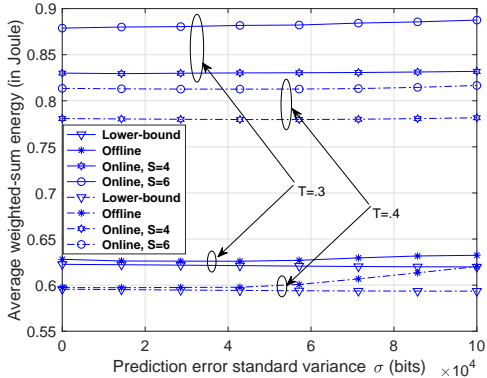}
	\caption{Average weighted-sum energy versus standard variance of the task-input length prediction error.}\label{fig:energy vs error std var}
	\vspace{-0.1in}
\end{figure}

Fig. \ref{fig:energy vs error std var} demonstrates the average weighted-sum energy versus the standard variance of the prediction error \(\Delta L_i\)'s. It is seen that our online algorithms under different deadline constraints are overall robust against a wide range of standard variance. In both cases of $T=0.3$ and $T=0.4$ seconds, the performance of the online scheme with a window length of  $S=6$ is  inferior to that with a window length of $S=4$ with noticeably larger gap in the more strict deadline constraint of $T=0.3$, which is due to the advantage of the short-size window in coping with uncertainties. Furthermore, the online algorithm with $S=6$ becomes worse off when \(\sigma\) exceeds about \(5\times10^4\)  (\(7\times10^4\)) in the case of $T=.3$ ($T=.4$) seconds, since the effectiveness of the long-term prediction starts being compromised by the increasing estimation error. 

\section{Conclusion}
This paper studied a one-UT one-AP fog computing system over a finite time-slotted horizon, in which each computation task was required to be executed by  the end of the slot, and dynamic computation caching was allowed such that the AP could decide whether to cache the current task results for relieving its computation burden in the future. Under the assumption of three-slot correlation and imperfect estimation of the task-input bit-length, a joint computation offloading and caching optimization problem was formulated to minimize the long-term weighted-sum energy consumption of the UT and the AP. To tackle the challenging mixed-integer non-convex problem, we approximated the problem by an SDP, based on which an offline solution assuming perfect knowledge of task-input length was provided, while a sliding-window based online scheme was also developed  to cater for unknown prediction error of the future task arrivals. By comparison with several benchmark schemes, the proposed online algorithm with short-size window demonstrated striking robustness against prediction error. In addition, the approximation was also shown to be near-optimal by numerical examples under practical settings. 

\begin{appendices}
\section{}\label{appendix:proof of rank-two for A}
Only  a sketch of the proof is provided herein due to the space limitation, and detailed proof will be presented in the longer version of this paper. First, by providing the (partial) Lagrangian of \(\mathrm{(P1^\prime)}\) in terms of \(\mv A^\ast\) and the associated Karush-Kuhn-Tucker (KKT) conditions, show that \(\mv A^\ast\in\mb{R}^{(N+1)\times(N+1)}\)  lies in the null space of a matrix containing a tri-diagonal sub-matrix. Next, show that under the above sufficient condition, he rank of this matrix is no less than $N-1$, and thus \({\rm rank}(\mv A^\ast)\le 2\) is proved.
\end{appendices}

\bibliographystyle{IEEEtran}
\bibliography{MEC_ref}

\end{document}